\documentclass[a4paper,11pt]{article}

\usepackage[a4paper,margin=2.5cm]{geometry}

\usepackage[utf8]{inputenc}

\usepackage[english]{babel}

\usepackage{amsfonts}

\usepackage{graphicx}

\usepackage{subfigure}

\usepackage[thinlines]{easytable}

\usepackage{natbib}

\setcitestyle{authoryear,round,semicolon}

\usepackage{hyperref}

\usepackage{algorithm}
\usepackage{algpseudocode}

\title{Forecasting with Neuro-Dynamic Programming}

\author{Pedro Afonso Fernandes\footnote{ORCID: \url{https://orcid.org/0000-0001-5762-5157}. Correspondence: Universidade Católica Portuguesa, Católica Lisbon School of Business \& Economics, Palma de Cima, Building 5, 4th floor, Room 5430, 1649-023 Lisboa, Portugal. Email: paf@ucp.pt. URL: \url{https://paf.com.pt}.}\\\\Universidade Católica Portuguesa\\Católica Lisbon School of Business \& Economics\\Católica Lisbon Research Unit in Business \& Economics (CUBE)\\Católica Lisbon Forecasting Lab (NECEP)\\Portugal}

\date{\selectlanguage{english} \today}

\begin{document}

\maketitle

\selectlanguage{english}

\begin{abstract}

Economic forecasting is concerned with the estimation of some variable like gross domestic product (GDP) in the next period given a set of variables that describes the current situation or state of the economy, including industrial production, retail trade turnover or economic confidence. Neuro-dynamic programming (NDP) provides tools to deal with forecasting and other sequential problems with such high-dimensional states spaces. Whereas conventional forecasting methods penalises the difference (or loss) between predicted and actual outcomes, NDP favours the difference between temporally successive predictions, following an interactive and trial-and-error approach. Past data provides a guidance to train the models, but in a different way from ordinary least squares (OLS) and other supervised learning methods, signalling the adjustment costs between sequential states. We found that it is possible to train a GDP forecasting model with data concerned with other countries that performs better than models trained with past data from the tested country (Portugal). In addition, we found that non-linear architectures to approximate the value function of a sequential problem, namely, neural networks can perform better than a simple linear architecture, lowering the out-of-sample mean absolute forecast error (MAE) by 32\% from an OLS model.\\\



\end{abstract}

\pagebreak


\pagebreak

\section{Introduction}
\label{sec:intro}

Economic forecasting is concerned with the estimation of some variable of interest like gross domestic product (GDP) in the next period given a set of variables that describes the current situation or state of the economy. This set can include known values of the variable of interest or other covariates that describe the environment faced by economic agents. Forecasting is then finding the optimal value of the parameters of some (typically) linear model by minimising a loss function that measures the difference between the observed and the predicted values of the variable of interest.

Thus, forecasting is concerned with sequential problems where some cost is minimised given the current and future states of the economy. \emph{Dynamic programming} (DP) is concerned with this kind of multistage decision processes that follow the \emph{principle of optimality}, that is, whatever the initial state and initial decision are, the remaining decisions must constitute an optimal policy with regard to the state resulting form the first decision \citep{Bellman1962}. Here, the \emph{state} is a finite-dimensional object that, from the point of view of current and future costs, completely summarises the current situation faced by economic agents or decision makers \cite[section 1.4]{Ljungqvist2018}. 

States must be kept in low dimension in order to solve optimally the functional (Bellman) equation of a dynamic problem. Nevertheless, DP has a wide range of applications, namely, in economics following the dynamic recursive theories formulated by \citet{Stockey1989} and \citet{Ljungqvist2018}. A good example is the savings problem where a representative household maximises the expected utility of the current and future consumption subject to a sequence of budget constraints that relates labour income, financial wealth and consumption. The state space of this model could be kept in a low discrete dimension, but complex models may require an approximation to the return function that expresses the optimal long run value of some state using, namely, quadratic forms, resulting in the so-called \emph{optimal linear regulator problem}. Even these models require point estimates of their parameters \citep{Mannor2007}, which constitute a practical limitation of DP, namely, in the field of economics.

Originally formulated by \citet{Bertsekas1996}, \emph{neuro-dynamic programming} (NDP), also known as \emph{approximate} DP or \emph{reinforcement learning} (in machine learning literature), provides a set of methods and tools to deal with high-dimensional state spaces and challenging DP problems. It can deal also with problems where a mathematical model is unavailable or hard to construct using model-free methods. Thus, NDP is primarily concerned with approximating the return function of a dynamic problem using linear or nonlinear functional forms. Within this framework, optimality is relaxed in order to solve complex and computationally intensive DP problems, where regular approaches are not applicable or are too costly \citep{Bertsekas2011}.

Simultaneously, NDP makes use of ideas from artificial intelligence involving simulation-based algorithms and neural networks \citep{vanRoy1997}. Reinforcement learning (RL) algorithms learn what to do so as to maximise some reward (or minimise some cost), having in mind a long-term objective \citep{Szepesvari2010}. The learner is not told which actions to take, but instead must discover which actions yield the most reward (or least cost) by trying them in an interactive and trial-and-error process \citep{Sutton2018}.

The objective of this paper is to test the viability of using NDP/RL methods and tools to approximate the GDP level with the state of the economy described by several variables, possibly using big data. These variables include readily available data that can be used to forecast GDP in the current quarter like industrial production, retail sales, exports or economic confidence. Past data concerned with GDP provides a guidance to train the models, but in a different way from ordinary least squares and other supervised learning methods, signalling the adjustment costs between sequential states. Our models were trained using quarterly data from European countries and tested in Portugal. We found that a neural network architecture with tensor product can perform quite well, namely, during COVID-19 times.

The paper is organised as follow: firstly, we provide a review of selected, relevant literature in section \ref{sec:literature}; then, in section \ref{sec:framework}, we describe succinctly two approaches to train NDP models using linear and nonlinear architectures, as well as the dynamic model here adopted to forecast GDP; a brief description of the data set and its treatment is made in section \ref{sec:data}; section \ref{sec:findings} presents the main findings, followed by a final conclusion in section \ref{sec:conclusion}.

\section{Literature}
\label{sec:literature}

This paper covers a topic with contributions from the fields of forecasting, economics, dynamic programming and reinforcement learning. Here we track only the most important references.

Dynamic programming has a wide range of applications in economics including the savings problem, economic growth, job search, business cycles, olipoly equilibrium or recursive contracts \citep{Ljungqvist2018}. Particularly important for the present research is the problem of a competitive firm that maximises the inter-temporal value of its production with an adjustment cost of the rate of output: in order to obtain a greater return (or value) in the future, the firm must invest (or set aside) a part of its current production incurring a quadratic cost, following a scheme like the one originally proposed by \citet{LucasJr.1971}. The trade-off between immediate costs and future returns gives the firm the incentive to forecast the output market price as far as the investment (or change of output) decision is concerned.  

This problem can be solved optimally by assuming that the firm's return function is quadratic on the state variables, namely, the current level of output. However, we have to know in advance the values of the model's parameters to get that kind of solution. Here, our goal is to estimate those parameters directly from data following the temporal difference (TD) method proposed by \citet{Sutton1988}. Whereas conventional forecasting methods penalises the difference (or loss) between predicted and actual outcomes, that method is guided by the difference between temporally successive predictions.   

Our goal is also concerned with using other architectures than linear-quadratic ones for the return function, namely, neural networks. The approximation of Bellman's value and policy functions with neural networks was originally proposed by \citet{Bertsekas1996} with lately developments described by \citet{Bertsekas2011, Bertsekas2020}. Two broad approaches are available in neuro-dynamic programming (NDP): the direct estimation of the parameters using temporal differences or its indirect estimation using a least squares scheme on a projected Bellman equation.

NDP applications include parking, maintenance and repair \citep{Bertsekas1996}, retailer inventory management \citep{vanRoy1997} and dynamic catalog mailing policies \citep{Mannor2007}. In addition, NDP is being applied with great success in machines that play masterly complex games like backgammon \citep{Tesauro1995}, Go \citep{Silver2016} or chess \citep{Thrun1995, McIlroy2020}. Most advanced algorithms attained a superhuman performance by \emph{tabula rasa} learning from games of self-play \citep{Silver2017}. Economic interpretations of algorithms that play board games can be found in \citet{Igami2020}.

The approximation of functions with neural networks is a recent research topic in economics with a seminal contribution in \citet{Duffy2001} who approximated the conditional expectation function in the Euler equation of a stochastic growth model. In fact, heterogeneous-agents models and other high-dimensional problems cannot be solved using classic dynamic programming due to the "curse of dimensionality" \citep{Bellman1962}. In this scope, \citet{Maliar2021} introduce a unified deep learning method that solves complex dynamic economic models by casting them into nonlinear regression equations. They use neural networks to perform model reduction and to handle multicollinearity. \citet{Kahou2021} propose a new method for solving high-dimensional dynamic programming problems and recursive competitive equilibria with a large number of heterogeneous agents using deep learning and exploring symmetry. \citet{Kase2022} take advantage of the scalability of neural networks to estimate nonlinear heterogeneous-agents models with likelihood methods. \citet{Azinovic2022} uses neural networks to compute approximate functional rational expectations equilibria of economic models featuring a significant amount of heterogeneity, uncertainty and binding restrictions, namely, a large-scale overlapping generations (OLG) model.    

In a broad sense, this paper is about the application of reinforcement learning (RL) methods to economic forecasting. RL is founded on learning from interaction with the environment \citep{Sutton2018}. Guided by an immediate reward (or cost), an agent may find an optimal long-run policy by exploit the most advantageous actions and explore new, potentially better actions. RL develops the optimal control (or DP) framework by introducing computer science techniques like deep neural networks. Complete surveys of RL applications in economics and finance can be found in \citet{Charpentier2021} and \citet{Meng2019}.

The use of NDP/RL to forecast GDP is a relatively new topic in literature, we few applications such the ensembling approach of \citet{Li2022} that combine three deep neural networks to forecast the regional GDP of China.

\section{Framework}
\label{sec:framework}

\subsection{Dynamic programming}

Dynamic programming is concerned with sequential problems where the state of a discrete-time dynamic system evolves according to given transition probabilities that depend on some control variable $u$ \citep[chapter 1]{Bertsekas1996}. Given an initial state $i$, we choose the control $u$ to minimise either the immediate cost $g(i,u,j)$ associated with the transition from $i$ to the next state $j$ or the optimal cost-to-go of $j$, denoted by $J^*(j)$, which is the expected cost of all subsequent periods starting from $j$. The objective of DP is to calculate the \emph{optimal cost-to-go function} $J^*$ by iterating on the following Bellman equation:

\begin{equation}
	J^*(i) = \min_{u} E \left[ g(i,u,j) + \alpha J^*(j) | i,u \right],
	\label{eq:bellman}
\end{equation}

\noindent for all $i$, where $\alpha \in (0,1)$ is a discount factor. The time-invariant rule $\mu(i)$ that maps each state $i$ to the optimal control $u$ that attains the minimum in equation (\ref{eq:bellman}) is called a \emph{policy} or \emph{feedback control policy}. It can be computed either simultaneously with the optimal cost-to-go $J^*$ or in real time by minimising the right-and-side of the Bellman equation.

\subsection{Approximate dynamic programming}

In many application, the number of states and controls becomes prohibitively large in terms of computing time. Neuro-dynamic  or approximate dynamic programming deals with this "curse of dimensionality" \citep{Bellman1962} by approximating the optimal cost-to-go function $J^*$ through a neural network or linear architecture $\tilde{J}(i,r)$, where $r$ is a parameter/weight vector \citep{Bertsekas1996, Bertsekas2011}. Once $r$ is determined, it yields a sub-optimal control at any state $i$ through the one-step look-ahead minimisation: 
 
\begin{equation}
	\tilde{\mu}(i) = \arg \min_{u} E \left[ g(i,u,j) + \alpha \tilde{J}(j,r) | i,u \right].
	\label{eq:aprox_bellman}
\end{equation}

The function $\tilde{J}$ is called the \emph{scoring function} or \emph{approximate cost-to-go function}, and the value $\tilde{J}(j,r)$ is called the \emph{score} or \emph{approximate cost-to-go} of state $j$ \citep{Bertsekas1996}. It is a \emph{compact representation} in the sense that it reduces the description of $j$ to a few relevant features whose weights are the parameters $r$. For example, in computer chess, the state is the current board position described by features like the midgame/endgame material point values, material imbalances, mobility and trapped pieces, pawn structure, king safety, outposts, bishop pair, and other evaluation patterns \citep{Bertsekas2011, Silver2017}.

\subsection{Neural network}
\label{sec:NN}

The scoring function $\tilde{J}$ can be approximated by a nonlinear architecture, namely, a \emph{feedforward neural network} (FFNN). Also know as \emph{deep neural network}, \emph{artificial neural network} or \emph{multilayer perceptron}, it is a simple model where several linear combinations of inputs are passed through nonlinear activation functions called \emph{nodes} \citep[chapter 10]{Taddy2019}. A set of nodes is called a \emph{layer}, and the output of ones layer's node becomes the input of the next layer in multilayer architectures. Finally, the output of the last layer is combined linearly to produce a cost-to-go.

For a state $i$ represented by a \emph{feature vector} $ x(i) = (x_1(i), \dots, x_q(i))'$, the first layer of a FFNN is composed by $s$ nodes $h_k(i)$ such that \citep[chapter 3]{Bertsekas1996}:

\begin{equation}
\label{eq:layer}
h_k(i) = \sigma \left(\sum_{l=1}^{q}{r_{kl}} x_l(i) \right), k = 1, \dots, s,
\end{equation}

\noindent where $\sigma(\cdot)$ is a nonlinear \emph{activation function} specified in advance. Early implementations of FFNN typically favoured the logistic function:

\begin{equation}
\label{eq:logistic}
\sigma(\xi) = \frac{1}{1 + e^{-\xi}} = \frac{e^{\xi}}{e^{\xi}+1}.
\end{equation}

\noindent Nowadays, the most common activation function is the ReLU (rectified linear unit):

\begin{equation}
\label{eq:relu}
\sigma(\xi) = \textrm{max} (0, \xi),
\end{equation}

\noindent because it can be computed and stored more efficiently than the logistic function and other options. The nonlinear nature of the activation function is fundamental to capture complex nonlinearities and interaction effects \citep[chapter 10]{James2021}.

The nodes given by equation (\ref{eq:layer}) can be used as input of a second layer whose output can be passed to a third layer, and so on. Following \citet[chapter 3]{Bertsekas1996}, we adopt a FFNN with a single layer whose final output is given by

\begin{equation}
\label{eq:output}
\tilde{J}(i,r) = \sum_{k=1}^{s}{r_k h_k(i)} = \sum_{k=1}^{s}{r_k \sigma \left(\sum_{l=1}^{q}{r_{kl}} x_l(i) \right)},
\end{equation}

\noindent where the parameter vector $r$ is composed by the weights $r_k$ and $r_{kl}$ for $k = 1, \dots, s$ and $l = 1, \dots, q$. These coefficients can be estimated from training data using \emph{temporal difference learning} (TD). Introduced by \citet{Sutton1988}, TD deals with multi-step prediction problems where a future outcome is predicted by updating the \emph{weights} of the model from the changes or \emph{errors} in successive predictions rather than from the overall error between predictions and the final outcome. Thus, our simple FFNN can be trained using the algorithm \ref{alg:TD0NN} \citep[section 2.2]{Szepesvari2010} \citep[section 6.2]{Bertsekas2011}:

\begin{algorithm}
\caption{TD(0) with nonlinear cost-to-go approximation (neural network)}
\label{alg:TD0NN}
\begin{algorithmic}[1]
\State $\delta \gets \tilde{J}(i,r) - \alpha \tilde{J}(j,r) - g(i,u,j)$
\State $r_k \gets r_k - \gamma \delta h_k(i)$
\State $r_{kl} \gets r_{kl} - \gamma \delta r_k \nabla_{r_{kl}}  h_k(i)$
\State \Return $r$
\end{algorithmic}
\end{algorithm}

\noindent where $\gamma$ is a small non-negative number, called \emph{step-size}, that decreases with time, and the gradient $\nabla_{r_{kl}}  h_k(i)$ is the partial derivative of the node $k$ with respect to the weight $r_{kl}$. Using the ReLU function (\ref{eq:relu}), this derivative is simply $x_l(i)$ if the weighted sum of features $\sum_{l=1}^{q}{r_{kl}} x_l(i)$ is positive and zero otherwise. The features $x_l(i)$ can be constructed from states individual components or attributes using the tensor product \citep[section 2.2]{Szepesvari2010}.

\subsection{Linear architecture}
\label{sec:LSTD}

Alternatively, the approximate cost-to-go function $\tilde{J}$ may adopt the linear form

\begin{equation}
\label{eq:linarch}
\tilde{J}(i,r) = \sum_{l=1}^{q}{r_l x_l(i)},
\end{equation}

\noindent where $r = (r_1, \dots, r_q)'$ and the features $x_l(i)$ are \emph{basis functions} because they form a linear basis of the set of approximate functions \citep[section 9.4]{Sutton2018}. The weights $r_l$ can be estimated with temporal differences using the algorithm \ref{alg:TD0LIN}.

\begin{algorithm}
\caption{TD(0) with linear cost-to-go approximation}
\label{alg:TD0LIN}
\begin{algorithmic}[1]
\State $\delta \gets \tilde{J}(i,r) - \alpha \tilde{J}(j,r) - g(i,u,j)$
\State $r_l \gets r_l - \gamma \delta x_l(i)$
\State \Return $r$
\end{algorithmic}
\end{algorithm}

\subsection{Dynamic model}
\label{sec:model}

In our simple model, output (GDP) is adjusted with a quadratic cost. The objective of a social planner is to minimise this immediate cost, as well as the future costs of adjustment. Thus, the cost-to-go function $J$ reflects the adjustment path of the economy as a function of several state variables that describe the economic environment. The Bellman equation of the model is 

\begin{equation}
	J(i) = \min_{u} E \left[ u^2 + \alpha J(j)\right],
	\label{eq:model}
\end{equation}

\noindent where $i$ is the state of the economy in the current quarter, $u$ is the quarter-over-quarter change of output and $j$ is the state in the next quarter. This model is basically a simplified version of the competitive firm model of \citet[section 7.2]{Ljungqvist2018} adapted from \citet{LucasJr.1971}.

\section{Data}
\label{sec:data}

The cost-to-go function $J$ in equation (\ref{eq:model}) was approximated using the two architectures and algorithms described in sections \ref{sec:NN} and \ref{sec:LSTD}. NDP/RL algorithms can converge slowly, thus requiring long batches with several observations to produce accurate approximations. Economic time series are typically too short for that purpose. Anyway, it is possible to consider multiple series concerned with several countries. Thus, we trained our models with a panel of 26 countries from the European Union, reserving for testing the data concerned with Portugal. Data covered the period from the first quarter of 2000 to the fourth quarter of 2023 with 95 transitions between adjacent quarters for each country, and a total of 2470 records in the training data set. 

Besides GDP, data covered the state variables indicated in table \ref{tab:states} that describe the economic environment in each country, covering production and turnover (industry, construction and services), international trade (exports and imports) and the confidence of economic agents measured by the Economic Sentiment Indicator (ESI). These variables are readily available and can be used to assess the current situation of the economy right before the dissemination of official macroeconomic aggregates from the national quarterly accounts (NQA), supporting a \emph{nowcasting} exercise \citep{Baffigi2004,Giannone2008}.     

\begin{table}[h!]
	\caption{List of state variables} 
	\label{tab:states} 
	\centering
	\begin{TAB}(@,2.5cm,0.5cm){lcc}{|c|cccccc|}
		Variable                              & Source    & Transformation\\
		Industrial production index           & Eurostat  & $(x-min)/(max-min)$\\
		Production in construction		      & Eurostat  & $(x-min)/(max-min)$\\
    	Retail trade turnover			      & Eurostat  & $(x-min)/(max-min)$\\
        Exports of goods                      & Eurostat  & $(x-min)/(max-min)$\\
        Imports of goods                      & Eurostat  & $(x-min)/(max-min)$\\
        Economic Sentiment Indicator (ESI)    & EC        & $(x-min)/(max-min)$\\       
    \end{TAB}
\end{table}

Each observation was regularised by subtracting the minimum value and dividing by the range, that is, the difference between the maximum and minimum values. This operation was performed for each indicator individually and for each country separately in order to assure the comparability between records. Regularisation is a critical procedure to estimate, namely, the neural network weights using the temporal difference algorithm \ref{alg:TD0NN}.

Data concerned with Portugal covers also the period 2000Q1-2023Q4. The observations up to 2014 (first 15 years) were used to train a benchmark model, that is, an ordinary least squares (OLS) regression of GDP on the variables listed in table \ref{tab:states}. Then, the predictive performance of the models was tested using the data for Portugal from the first quarter of 2015 to the fourth quarter of 2023.

\section{Findings}
\label{sec:findings}

Table \ref{tab:findings} presents the out-of-sample (OOS) mean absolute error (MAE) and root mean squared error (RMSE) for the models previously described, that is, a linear regression (benchmark), a neural network and a linear architecture, both estimated with temporal differences. The response was the regularised Portuguese GDP and the OOS errors were calculated for the testing period between the first quarter of 2015 and the fourth quarter of 2023.

\begin{table}[h!]
	\caption{OOS errors by model (regularised GDP level, Portugal, 2015Q1-2023Q4)} 
	\label{tab:findings} 
	\centering
	\begin{tabular}{lcc}
		  \hline
        Model                           & MAE     & RMSE   \\
	      \hline	
        Linear regression (OLS)         & 0.0661  & 0.0832 \\
		TD(0) with neural network       & 0.0453  & 0.0597 \\
        TD(0) with linear architecture  & 0.1070  & 0.1281 \\
        \hline
    \end{tabular}
\end{table}

Firstly, we found that an OLS model of GDP level on industrial production, construction output, retail trade, international trade and economic sentiment can perform quite well. In fact, this model provides a MAE of only 6.6\% of the range between the maximum and minimum levels of GDP for the testing period. The RMSE is about 8.3\% of that range.

\begin{figure}[!h]
	\centering
	\includegraphics[width=14cm]{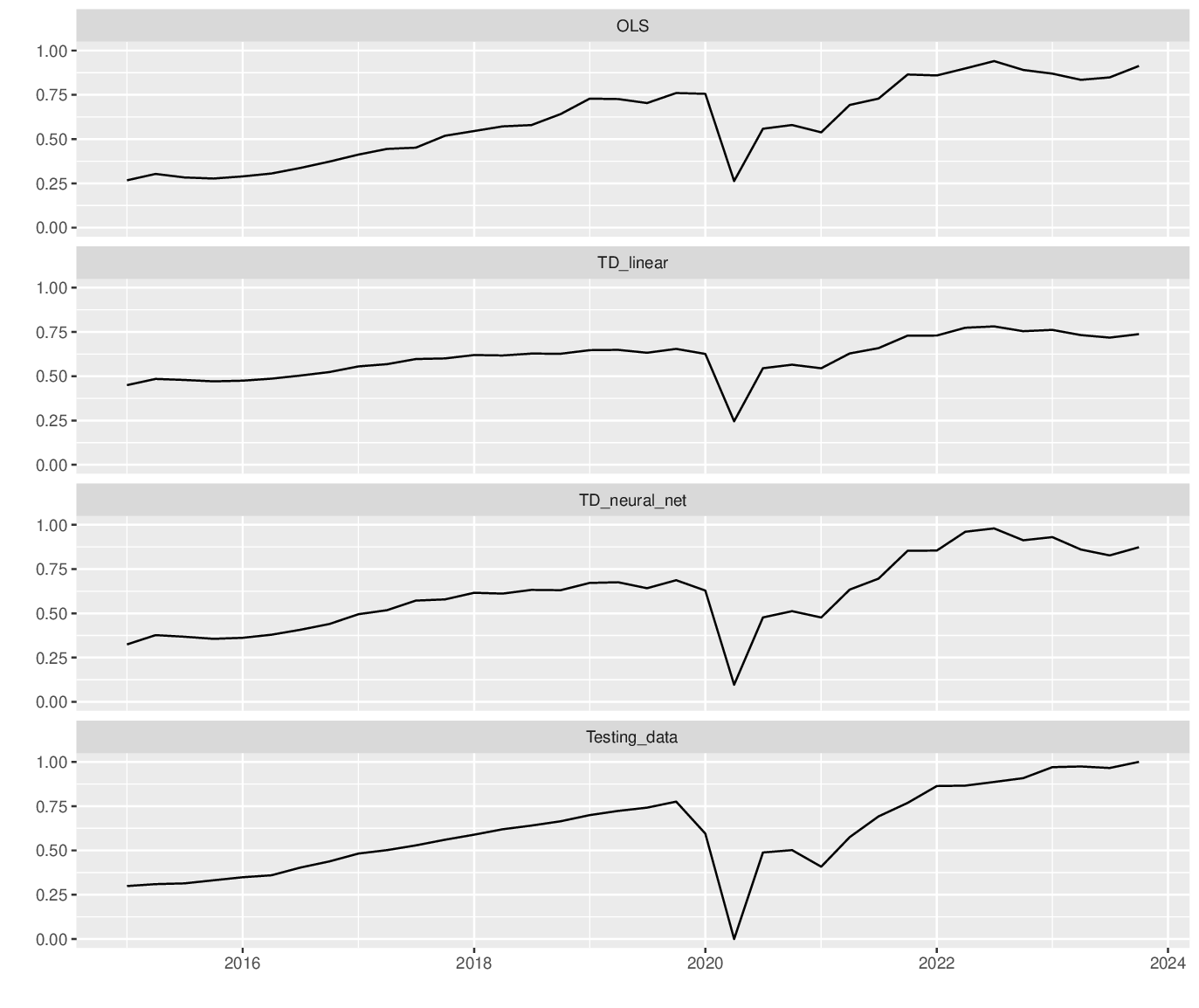}
	\caption{Forecasts of the regularised GDP level by different models and testing data (Portugal, 2015Q1-2023Q4)}
	\label{fig:NDP}
\end{figure}

\begin{figure}[!h]
	\centering
	\includegraphics[width=14cm]{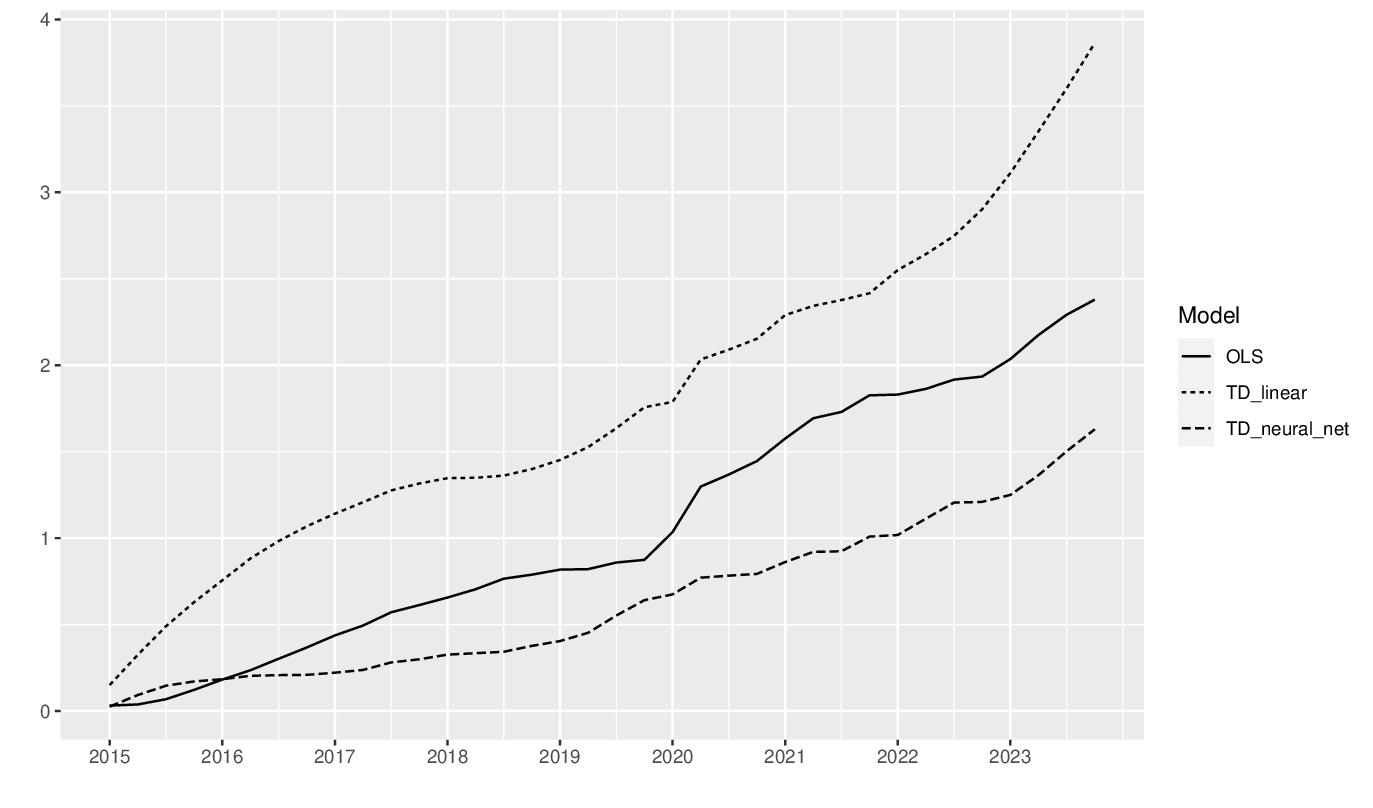}
	\caption{Cumulative absolute OOS error of the regularised GDP level by model  (Portugal, 2015Q1-2023Q4)}
	\label{fig:error}
\end{figure}

Secondly, a neural network trained with data for the European Union countries (excluding Portugal) can perform even better, lowering that MAE by 32\% to 4.5\% and the RMSE by 28\% to 5.9\%. In fact, the non-linear nature of this architecture might capture well the dramatic drop in GDP occurred during the "Great Lockdown" of the spring of 2020, as suggested by figure \ref{fig:NDP}, lowering the OOS errors for the testing period. In addition, figure \ref{fig:error} reveals that TD with neural network was quite effective during the COVID-19 times (since 2020) in mitigating the cumulative absolute OOS error.

Besides, the neural network performs better than a linear architecture estimated with temporal differences too. The last model is the worst with a MAE of 10.7\% of the range between the maximum and minimum levels of GDP, and a RMSE of 12.8\% of that range. Thus, a simple least squares model estimated with a restricted dataset (for Portugal) can perform better than a linear model estimated with TD over a full dataset (for the EU countries except Portugal), but so well than a neural network trained with the last dataset.

\section{Conclusion}
\label{sec:conclusion}

NDP/RL algorithms proved to achieve remarkable results using generic or even simulated data, not directly connected with the predicted object, namely in the field of board games. Here, we found that it is possible to train a model with data concerned with other countries that performs quite well, even better than models trained with past data from the tested country. In fact, time series could be too short to train effectively forecasting models, and some advantage may be obtained by using panel data from other countries with  algorithms that learn by trial-and-error, guided by the change of output as an adjustment cost. 

In addition, we found that non-linear architectures to approximate the value function of a sequential problem, namely, neural networks with tensor product can perform better than a simple linear architecture as far as the prediction of GDP is concerned, despite the good achievement obtained by a linear model estimated with ordinary least squares using past data for the region of interest.

\section*{Acknowledgments}
\label{sec:acknowledgments}

This work was supported by Fundação para a Ciência e Tecnologia (FCT), Lisbon, Portugal under a doctorate auxiliary researcher grant with the reference CUBE-PhD-CEEC/1 from Católica Lisbon Research Unit in Business \& Economics (UID/GES/00407/2020).

\section*{Abbreviations}
\label{sec:abbreviations}

The following abbreviations are used in this paper:\\

\noindent 
\begin{tabular}{@{}ll}
DP      & Dynamic Programming\\
EC      & European Commission\\
ESI     & Economic Sentiment Indicator\\
FCT     & Portuguese Foundation for Science and Technology\\
FFNN    & Feedforward Neural Network\\
GDP     & Gross Domestic Product\\
MAE     & Mean Absolute Error\\
NDP     & Neuro-Dynamic Programming\\
NQA     & National Quarterly Accounts\\
OLG     & Overlapping Generations\\
OLS     & Ordinary Least Squares\\
OOS     & Out-Of-Sample\\
ReLU    & Rectified Linear Unit\\
RL      & Reinforcement Learning\\
RMSE    & Root Mean Squared Error\\
TD      & Temporal Difference
\end{tabular}

\bibliography{library}

\end{document}